\documentclass[conference]{IEEEtran}
\IEEEoverridecommandlockouts

\usepackage{cite}
\usepackage{amsmath,amssymb,amsfonts}
\usepackage{algorithmic}
\usepackage{graphicx}
\usepackage{textcomp}
\usepackage{xcolor}
\def\BibTeX{{\rm B\kern-.05em{\sc i\kern-.025em b}\kern-.08em
    T\kern-.1667em\lower.7ex\hbox{E}\kern-.125emX}}
\usepackage{braket}

\begin{document}

\title{On Quantum BSDE Solver for High-Dimensional Parabolic PDEs}

\author{\IEEEauthorblockN{Howard Su}
\IEEEauthorblockA{\textit{Department of Computing} \\
\textit{Imperial College London}\\
London, UK \\
h.su24@imperial.ac.uk}
\and
\IEEEauthorblockN{Huan-Hsin Tseng}
\IEEEauthorblockA{\textit{AI \& ML Department} \\
\textit{Brookhaven National Laboratory}\\
NY, USA \\
htseng@bnl.gov}
}

\maketitle

\begin{abstract}

We propose a quantum machine learning framework for approximating solutions to high-dimensional parabolic partial differential equations (PDEs) that can be reformulated as backward stochastic differential equations (BSDEs). In contrast to popular quantum-classical network hybrid approaches, this study employs the pure Variational Quantum Circuit (VQC) as the core solver without trainable classical neural networks.

The quantum BSDE solver performs pathwise approximation via temporal discretization and Monte Carlo simulation, framed as model-based reinforcement learning. We benchmark VQC-based and classical deep neural network (DNN) solvers on two canonical PDEs as representatives: the Black–Scholes and nonlinear Hamilton–Jacobi–Bellman (HJB) equations. The VQC achieves lower variance and improved accuracy in most cases, particularly in highly nonlinear regimes and for out-of-the-money options, demonstrating greater robustness than DNNs. These results, obtained via quantum circuit simulation, highlight the potential of VQCs as scalable and stable solvers for high-dimensional stochastic control problems.
\end{abstract}


\section{Introduction}

High-dimensional parabolic partial differential equations (PDEs) arise in a wide range of domains, including physics, quantitative finance, and stochastic control. However, numerically solving such equations remains a significant challenge. Classical grid-based numerical methods, such as finite difference and finite element schemes, suffer from the curse of dimensionality, with computational cost growing exponentially in the number of spatial dimensions. As a result, these methods are typically limited to problems with at most three or four dimensions.

In recent years, deep learning-based approaches have emerged as a promising alternative for overcoming this bottleneck. Techniques such as Physics-Informed Neural Networks (PINNs), the Deep Galerkin Method (DGM), and BSDE-based methods aim to approximate PDE solutions through neural networks trained on data or stochastic representations. We refer the reader to the comprehensive reviews of Beck et al.~\cite{beck2022overview} and Blechschmidt and Ernst~\cite{blechschmidt2021three}.

In this work, we adopt the BSDE-based (backward stochastic differential equations) approach due to its flexibility and generality. The BSDE formulation provides a probabilistic representation that applies to linear, semilinear, and fully nonlinear PDEs. It naturally scales to high-dimensional settings and admits solutions over the full time-space domain. Furthermore, BSDEs are closely linked to stochastic control and optimal stopping problems, making them particularly well-suited for applications in quantitative finance. Importantly, the BSDE solver can be viewed as a model-based reinforcement learning algorithm: the BSDE itself acts as a model of the system, the gradient of the solution serves as a policy function, and the terminal loss defines the task objective. This perspective parallels deep reinforcement learning, where neural networks approximate policies or value functions within a learned model of the environment, as discussed by E, Han, and Jentzen~\cite{e2017deep}.

The mathematical foundation of this approach is rooted in the seminal work of Pardoux and Peng~\cite{pardoux1992backward}, who introduced the theory of BSDEs and established that a broad class of quasilinear parabolic PDEs can be represented probabilistically through nonlinear Feynman-Kac formulas. This representation expresses the PDE solution as a conditional expectation governed by a backward stochastic system, enabling Monte Carlo-based solution methods for high-dimensional problems. This framework was later extended to forward-backward stochastic differential equations (FBSDEs) by Pardoux and Tang~\cite{pardoux1999forward}, further deepening the connection between stochastic processes and quasilinear PDEs.

Building on this foundation, the deep BSDE method proposed by E, Han, and Jentzen~\cite{e2017deep}, and further developed in~\cite{han2018solving}, uses DNNs to approximate the adapted pair of processes $(Y_t, Z_t)$ in the BSDE. The $Y_t$ component approximates the PDE solution, while the $Z_t$ component, interpreted as the spatial gradient, plays a critical role in learning dynamics and policy estimation. This hybrid of stochastic analysis and machine learning provides a scalable framework for solving complex PDEs and optimal control problems.

Advances in quantum computing have introduced promising tools for numerical scientific computing, notably through the development of \emph{variational quantum algorithms}~\cite{cerezo2021variational}. These hybrid methods combine parameterised quantum circuits with classical optimisation and are well-suited for noisy-intermediate-scale quantum (NISQ) devices. Key examples include the Variational Quantum Eigensolver (VQE)~\cite{peruzzo2014variational, mcclean2016theory} for electronic-structure problems and the Quantum Approximate Optimisation Algorithm (QAOA)~\cite{farhi2014quantum, zhou2020qaoa} for combinatorial optimisation.

Within this framework, \emph{variational quantum circuits} (VQCs)~\cite{mitarai2018quantum} serve as quantum analogues of neural networks, encoding classical data into quantum states, applying trainable unitaries, and extracting features via expectation measurements. Their expressiveness scales with circuit size, allowing versatile modelling. Beyond early success in supervised learning and function approximation, VQCs have also performed well in deep reinforcement learning~\cite{chen2020variational} and sequence modelling with quantum LSTM architectures~\cite{chen2022quantum}, supporting their potential as drop-in replacements for classical networks in BSDE solvers.

In this work, we propose a \emph{pure} VQC-based BSDE solver, where all trainable parameters are only in VQCs with randomly initialized and fixed classical layers serving as dimensional adapters. We benchmark the method on two representative parabolic PDEs: the Black-Scholes equation and a nonlinear Hamilton-Jacobi-Bellman (HJB) equation. Both examples admit closed-form analytical solutions, enabling rigorous error assessment. The Black-Scholes equation serves as a classical linear instance from quantitative finance, whereas the HJB equation introduces strong nonlinearities through its gradient dependence and represents an optimal control problem. These case studies allow us to systematically compare VQC-based and classical DNN-based solvers and to demonstrate the feasibility of pure VQC architectures for parabolic PDE frameworks in high-dimensional settings.

\section{BSDE Framework for General Parabolic PDEs}\label{Sec. BSDE framework}

\subsection{Problem Formulation}

Let $T > 0$ be a fixed time horizon and $d \in \mathbb{N}$ be the spatial dimension. We consider a general semilinear parabolic PDE of the form,
\begin{equation} \label{eq:pde}
\begin{aligned}
&\frac{\partial u}{\partial t}(t, x) + \mathcal{L} u(t, x) + f(t, x, u(t,x), \nabla_x u(t,x)) = 0,\\
& u(T, x) = g(x)
\end{aligned}
\end{equation}
where $(t, x) \in [0,T) \times \mathbb{R}^d$, $u : [0,T] \times \mathbb{R}^d \to \mathbb{R}$ is the unknown function to be solved, $g : \mathbb{R}^d \to \mathbb{R}$ is a prescribed terminal condition, $f : [0,T] \times \mathbb{R}^d \times \mathbb{R} \times \mathbb{R}^d \to \mathbb{R}$ is a given driver (possibly nonlinear), and $\mathcal{L}$ is a second-order differential operator of the form,
\begin{equation}
\label{eq:generator}
\mathcal{L} u(t,x) = \frac{1}{2} \text{Tr} \left(\sigma(x) \sigma (x)^\top \nabla_x^2 u(t,x) \right) + b(x)^\top \nabla_x u(t,x),
\end{equation}
where $b : \mathbb{R}^d \to \mathbb{R}^d$ is a drift vector field, and $\sigma : \mathbb{R}^d \to \mathbb{R}^{d \times d}$ is a diffusion matrix. 

We make the following regularity assumptions:

\begin{itemize}
    \item The functions $b$, $\sigma$, $f$, and $g$ are measurable, and there exists $C > 0$ such that
    \[
    |b(x)| + \|\sigma(x)\| + |f(t,x,y,z)| + |g(x)| \leq C (1 + |x| + |y| + |z|),
    \]
    for all $(t,x,y,z) \in [0,T] \times \mathbb{R}^d \times \mathbb{R} \times \mathbb{R}^d$.
    \item The function $f(t, x, y, z)$ is Lipschitz continuous in $(y,z)$ uniformly in $(t,x)$.
\end{itemize}

Under these conditions, it is known that the PDE \eqref{eq:pde} admits a unique classical solution under suitable additional smoothness assumptions (see, e.g.,~\cite{pardoux1992backward, pardoux1999forward}).

\subsection{Probabilistic Representation via Nonlinear Feynman-Kac Formula}
We now describe the probabilistic representation of the PDE solution using the nonlinear Feynman-Kac formula~\cite{pardoux1992backward}.

Let $X_t$ be the solution to the forward stochastic differential equation (SDE):
\begin{equation}
\label{eq:sde-forward}
dX_t = b(X_t) \, dt + \sigma(X_t) \, dW_t, \quad X_0 = \xi,
\end{equation}
where $(W_t)_{t \in [0,T]}$ is a $d$-dimensional standard Brownian motion defined on a filtered probability space $(\Omega, \mathcal{F}, (\mathcal{F}_t)_{t \in [0,T]}, \mathbb{P})$.

We associate with Eq.~\eqref{eq:sde-forward} the backward SDE:
\begin{equation}
\label{eq:bsde}
dY_t = -f(t, X_t, Y_t, Z_t) \, dt + Z_t^\top dW_t, \quad Y_T = g(X_T).
\end{equation}

Under suitable conditions, the solution $u(t,x)$ of the PDE \eqref{eq:pde} can be represented as
\begin{equation}
\label{eq:feynman-kac}
u(t,x) = \mathbb{E}[Y_t \mid X_t = x].
\end{equation}
Moreover, the spatial gradient satisfies:
\begin{equation}
\label{eq:gradient-relation}
\nabla_x u(t,x) = (\sigma(x)^\top)^{-1} \mathbb{E}[Z_t \mid X_t = x].
\end{equation}

\paragraph{Sketch of Derivation.} Applying It\^o's formula to $u(t,X_t)$ gives:
\begin{equation}
du(t, X_t) = \left(\frac{\partial u}{\partial t}(t, X_t) + \mathcal{L} u(t, X_t)\right)dt + \nabla_x u(t, X_t)^\top \sigma(X_t) dW_t.
\end{equation}
Using the PDE \eqref{eq:pde}, we identify the dynamics with Eq.~\eqref{eq:bsde} by setting:
\begin{equation}
Y_t = u(t, X_t), \quad Z_t = \sigma(X_t)^\top \nabla_x u(t, X_t).
\end{equation}

\subsection{Forward--Backward Stochastic System Representation}
Let \( \xi \in \mathbb{R}^d \) denote the initial condition of the forward process. We summarize the coupled system:
\begin{align}
&dX_t = b(X_t) dt + \sigma(X_t) dW_t, \quad X_0 = \xi, \\
&dY_t = -f(t, X_t, Y_t, Z_t) dt + Z_t^\top dW_t, \quad Y_T = g(X_T),
\end{align}
with $Z_t = \sigma(X_t)^\top \nabla_x u(t, X_t)$. This forward-backward system provides the foundation for our solver architecture, where the forward SDE governs the sample paths and the backward SDE encodes the PDE solution.

\subsection{Stochastic Control Interpretation}
The BSDE \eqref{eq:bsde} can be interpreted as a stochastic control problem. Consider:
\begin{equation}
Y^{y,Z}_t = y - \int_0^t f(s, X_s, Y^{y,Z}_s, Z_s) ds + \int_0^t Z_s^\top dW_s.
\end{equation}
The optimal control problem seeks:
\begin{equation}
\inf_{(y, Z)} \mathbb{E}\left[\left|Y_T^{y,Z} - g(X_T)\right|^2\right].
\end{equation}
It can be shown \cite{pardoux1992backward} that the minimizer satisfies $y^* = u(0, \xi)$ and $Z^*_t = \sigma(X_t)^\top \nabla_x u(t, X_t)$.

\subsection{Numerical Discretization}

For a numerical approximation of solution \eqref{eq:bsde}, we discretize time interval $[0,T]$ using a uniform grid:
\[
0 = t_0 < t_1 < \cdots < t_N = T, \quad \Delta t = \frac{T}{N}.
\]

We approximate the forward process $X_t$ using the Euler-Maruyama scheme:
\begin{equation}\label{eq:euler_maruyama}
X_{t_{n+1}} = X_{t_n} + b(X_{t_n}) \, \Delta t + \sigma(X_{t_n}) \, \Delta W_{t_n},
\end{equation}

where $\Delta W_{t_n} \sim \mathcal{N}(0, \Delta t \cdot I_d)$ represents the Brownian increment. 

Similarly, we discretize the BSDE backward in time:
\begin{equation}
\label{eq:bsde_discrete}
Y_{t_{n}} = Y_{t_{n+1}} + f(t_{n+1}, X_{t_{n+1}}, Y_{t_{n+1}}, Z_{t_{n+1}}) \, \Delta t - Z_{t_{n+1}}^\top \, \Delta W_{t_n}.
\end{equation}

The terminal condition is imposed at $t_N = T$:
\[
Y_{t_N} = g(X_{t_N}).
\]

The key computational task is to approximate the sequence of control variables $\{Z_{t_n}\}_{n=0}^{N-1}$. In this work, we explore using VQCs to parameterize the map $(t_n, X_{t_n}) \mapsto (t_n, Z_{t_n})$, fully replacing the DNN traditionally used in deep BSDE methods.

\subsection{Training Objective}

Finally, the entire system is trained by minimizing the expected squared terminal loss:
\begin{equation}
\label{eq:training_loss}
L(\theta) = \mathbb{E} \left[ \left| Y^\theta_{T} - g(X_{T}) \right|^2 \right],
\end{equation}
where $\theta$ denotes the parameters of the function approximator (in our case, VQC) used to represent $Z_{t_n}$. The expectation is estimated using Monte Carlo simulation.

\section{Quantum Machine Learning via VQC}
\subsection{General Framework of VQC}\label{Subsec: VQC framework}

VQCs are parameterized quantum circuits to learn data and signals, forming a core component of QML. The detailed construction is as follows. Let $\mathcal{H}=(\mathbb{C}^2)^{\otimes n}\cong\mathbb{C}^{2^n}$ be the $n$-qubit Hilbert space and $\mathcal{U}(\mathcal{H})$ its unitary group.  For input $x\in\mathbb{R}^n$ with label $y\in\mathbb{R}^k$, a VQC has following steps:

\paragraph{Encoding}
A data‐dependent unitary $V(x)\in\mathcal{U}(\mathcal{H})$ maps a reference state to
\begin{equation}
\ket{\psi_x}
=V(x)\ket{\psi_0}\,,\qquad \ket{\psi_0}\in\mathcal{H}.
\end{equation}

\paragraph{Parameterized evolution}
Using Pauli matrices $\sigma_1,\sigma_2,\sigma_3$, define a layered unitary
\begin{equation}
U(\theta) =\prod_{\ell=1}^L
\Bigl(\!\bigotimes_{q=1}^n e^{-\,\tfrac{i}{2}\,\theta_q^{(\ell)}\sigma_q}\Bigr)\,
\mathcal{C}_\ell \, \in\mathcal{U}(\mathcal{H}),
\end{equation}
which acts on $\ket{\psi_x}$ to produce $U(\theta)\ket{\psi_x}$.

\paragraph{Measurement \& output}
Choose Hermitian operators as observables $\{H_j\}_{j=1}^k$. The VQC (model) output is
\begin{equation}
f_{\mathrm{VQC},\theta}(x)
=\bigl(\langle H_1\rangle(x),\dots,\langle H_k\rangle(x)\bigr),
\end{equation}
with $\langle H_j\rangle(x) := \bra{\psi_0}V^\dagger(x)\,U^\dagger(\theta)\,H_j\,U(\theta)\,V(x)\ket{\psi_0}.$

\begin{figure}[htbp]
\vskip -0.15in
\begin{center}
\centerline{\includegraphics[width=0.9\columnwidth]{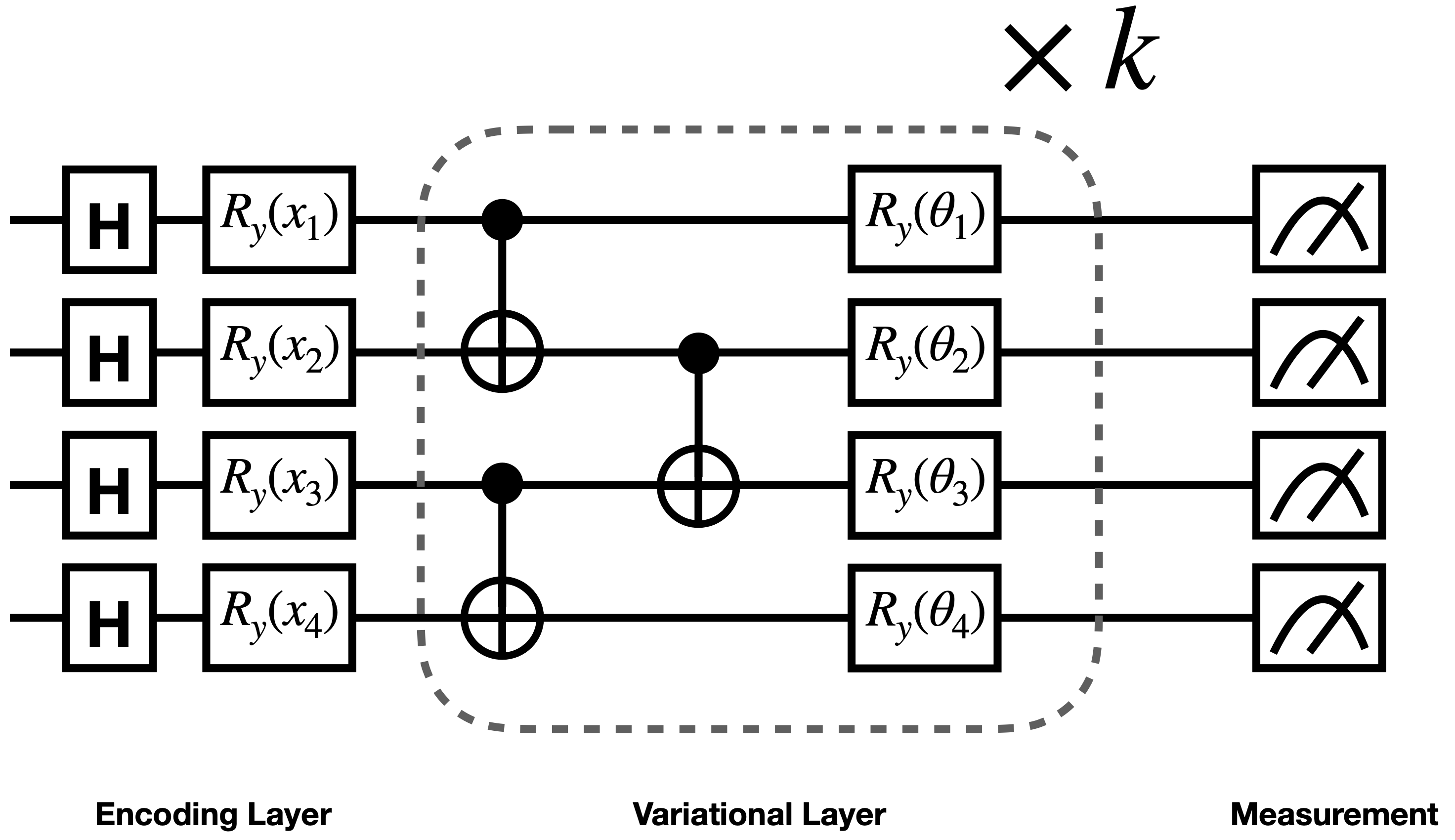}}
\caption{Schematic diagram of a VQC.}
    \label{fig:VQC}
\end{center}
\vskip -0.1in
\end{figure}

Fig.~\ref{fig:VQC} provides a schematic overview of these stages.

\begin{figure}[htbp]
\vskip -0.15in
\begin{center}

\centerline{\includegraphics[width=1.0\columnwidth]{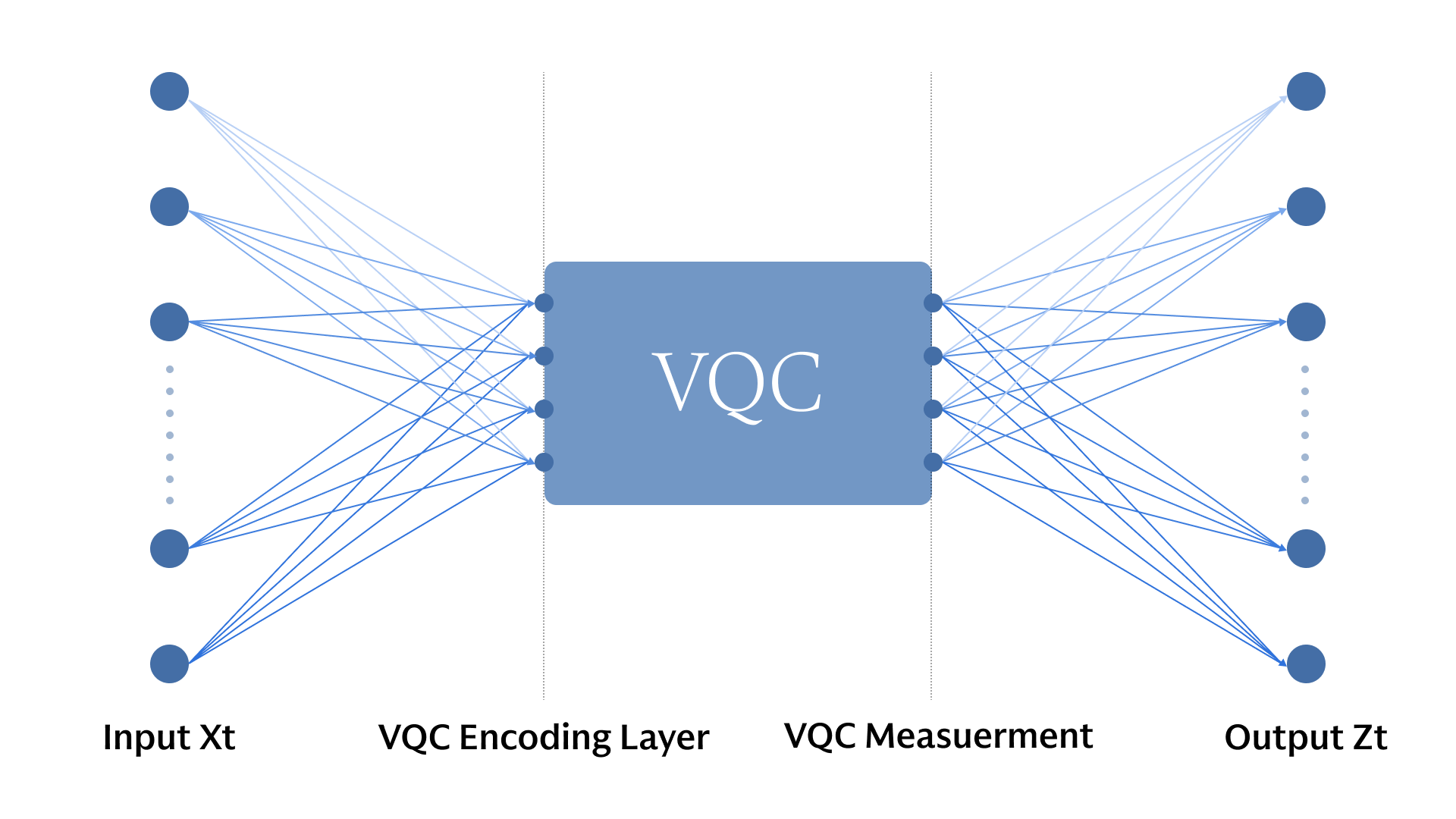}}
\caption{Structure of the VQC-based solver network}
    \label{fig:vqc_network}
\end{center}
\vskip -0.1in
\end{figure}

\subsection{Implementation Details of the VQC-Based BSDE Solver}
Building on the VQC framework in Sec.~\ref{Subsec: VQC framework}, our BSDE solver mainly relies on a \textit{purely quantum} architecture, where all \textit{trainable parameters} reside in the variational circuits. Due to the high dimensions of the BSDEs, it requires more qubits than what current VQCs can be implemented. A randomly initialized classical encoder–decoder pair is placed to reduce and restore the input dimension and kept fixed throughout training. These classical components serve only as adapters and do not participate in learning. \textit{Figure~\ref{fig:vqc_network}} illustrates this architecture: the high‑dimensional input is compressed by the encoder to match the number of qubits, processed by the VQC, and then expanded back to the original dimension by the decoder.

\paragraph{Forward SDE \(X_t\)}
We generate the forward trajectory \(X_t\) using the Euler–Maruyama scheme described in Eq.~\eqref{eq:euler_maruyama}. The simulated values serve as the input to the encoder at each time step.

\paragraph{Fixed Classical Encoding}
High-dimensional inputs are transformed via a fixed, randomly initialized classical linear layer $\mathbb{R}^d \to \mathbb{R}^{n_{\text{qubits}}}$ as encoding ($n_{\text{qubits}} \ll d$). This mapping is non‑trainable only to reduce data dimensions. 

\paragraph{Quantum Processing via VQC}
The dimension-reduced input is subsequently processed by a parameterized quantum circuit, which contains all trainable parameters such that the VQC learns the control process by mapping the current time and state \((t, X_t)\) to a control pair \((t, Z_t)\) required by the BSDE formulation.

\paragraph{Fixed Classical Decoding}
The VQC output is mapped back to the original dimension through another classical linear layer $\mathbb{R}^{n_{\text{qubits}}} \to \mathbb{R}^{d}$ fixed.

\paragraph{Backward SDE \(Y_t\)}
Once the decoder provides the control process \(Z_t\), we simulate the BSDE backward in time using the discretised update from Eq.~\eqref{eq:bsde_discrete}. Starting from the terminal condition \(Y_T = g(X_T)\), we recursively compute \(Y_t\) down to \(Y_0\). Throughout this backward recursion, the learned control \(Z_t\) couples into the BSDE dynamics.

\paragraph{Training Objective}
A classical Adam optimizer is used to minimize the discrepancy between the simulated terminal value and the known terminal condition. Specifically, the expected squared error between \( Y_T^\theta \) and the prescribed terminal payoff \( g(X_T) \) is minimized:
\[
L(\theta) = \mathbb{E} \left[ \left| Y_T^\theta - g(X_T) \right|^2 \right].
\]
This objective is used to train the quantum parameters $\theta$ within the VQC, enabling the model to learn the control strategy $Z_t$ that minimizes the discrepancy between the simulated and target terminal condition.

\section{Numerical Examples: Linear and Nonlinear Parabolic PDEs}

In this section, we apply the quantum BSDE framework described in Sec.~\ref{Sec. BSDE framework} to two canonical examples of parabolic PDEs:

\begin{itemize}
    \item the linear Black-Scholes equation from quantitative finance, and
    \item a nonlinear HJB equation arising in stochastic optimal control.
\end{itemize}

Both PDEs admit exact analytical solutions, which allow us to rigorously benchmark the accuracy of the learned approximations. In each case, we compare the performance of two approximator architectures for the control process $Z_t$: a classical DNN, and a VQC.

\subsection{Black--Scholes Equation}

\subsubsection{PDE Formulation}

Under the risk-neutral pricing framework, the value function $u(t,x)$ of a European option satisfies the following linear parabolic PDE:
\begin{equation}
\label{eq:black_scholes_pde}
\begin{split}
&\frac{\partial u}{\partial t}(t,x) + r x \frac{\partial u}{\partial x}(t,x) + \frac{1}{2} \sigma^2 x^2 \frac{\partial^2 u}{\partial x^2}(t,x) - r u(t,x) = 0,\\
&(t,x) \in [0,T) \times (0, \infty),
\end{split}
\end{equation}
with terminal condition:
\begin{equation}
\label{eq:black_scholes_terminal}
u(T,x) = g(x) = \max(\pm(x - K), 0),
\end{equation}
where $+$ corresponds to a call option and $-$ to a put option. 

Here $r > 0$ is the risk-free interest rate, $\sigma > 0$ is the volatility, and $K > 0$ is the strike price.

\subsubsection{BSDE Representation}

The Black--Scholes PDE corresponds to the following forward SDE for the asset price $X_t$:
\begin{equation}
\label{eq:bs_forward_sde}
dX_t = r X_t \, dt + \sigma X_t \, dW_t, \quad X_0 = x_0 > 0.
\end{equation}
and applying It\^o's lemma yields:
\[
d(\log X_t) = \left( r - \frac{1}{2} \sigma^2 \right) dt + \sigma dW_t.
\]
For numerical stability, we perform the simulation in logarithmic variables. We thus simulate $\log X_t$ directly.

The corresponding BSDE representation is:
\begin{equation}\label{eq:bs_bsde}
\begin{aligned}
dY_t &= - r Y_t \, dt + Z_t^\top \, dW_t, \quad t \in [0,T), \\
Y_T &= g(X_T).
\end{aligned}
\end{equation}

\subsubsection{Numerical Setup}

We consider 16 distinct portfolios, each treated as an independent testing case. Each portfolio consists of 100 European options of the same type, either calls or puts, with a fixed strike price. The strike prices span the range $K \in \{70,80,90,100,110,120,130,140\}$, resulting in 8 call portfolios and 8 put portfolios. This setup enables us to evaluate solver performance systematically across varying levels of moneyness in a 100-dimensional setting, where each input path includes 100 option payoffs, making the problem computationally intensive and reflective of realistic high-dimensional derivative pricing scenarios.

The model parameters are:
\[
r = 0.1, \quad \sigma = 0.2, \quad X_0 = 100, \quad T = 1~\text{year}.
\]
We discretize time into $N = 10$ steps of size $\Delta t = T/N$, and use $10,000$ Monte Carlo sample paths.

We compare two architectures for approximating $Z_t$:

\begin{itemize}
    \item DNN: 4 hidden layers, 64 neurons per layer, ReLU activation, batch size 100, learning rate 0.01, 10 epochs.
    \item VQC: 4 qubits, 2 VQC layers, batch size 100, learning rate 0.01, 10 epochs.
\end{itemize}

The training objective is to minimize the expected terminal squared loss:
\[
L(\theta) = \mathbb{E} \left[ \left| Y^\theta_T - g(X_T) \right|^2 \right].
\]

\begin{figure}[ht]
    \centering
\includegraphics[width=1\linewidth]{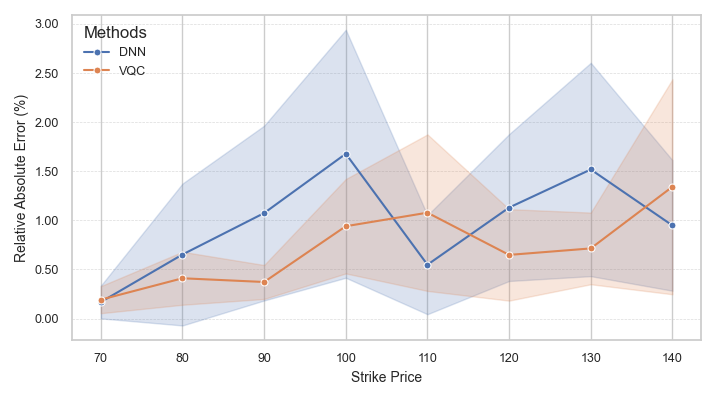} 
    \caption{Relative error (\%) for VQC and DNN across option strikes in the call option portfolio.}
    \label{fig:black_scholes_CALL_results}
\end{figure}

\begin{figure}[h]
    \centering
\includegraphics[width=1\linewidth]{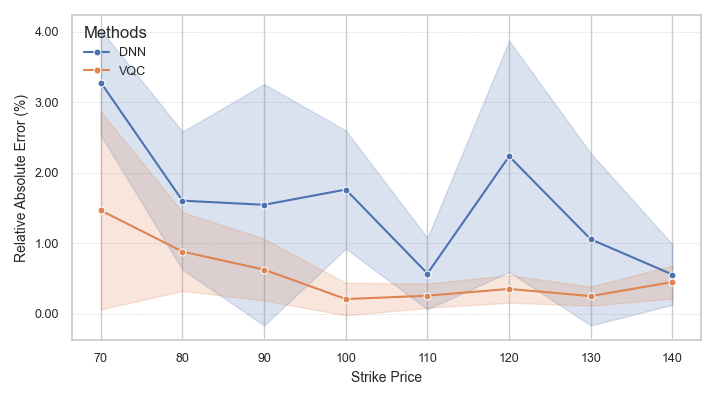} 
    \caption{Relative error (\%) for VQC and DNN across option strikes in the put option portfolio.}
    \label{fig:black_scholes_PUT_results}
\end{figure}

\subsubsection{Results and Discussion}

We assess the performance of the DNN and VQC solvers on a portfolio of 100 European options. The pricing results are compared against the analytical Black–Scholes solution, which is considered the true value. Each experiment is repeated five times, and we report the mean and standard deviation of the relative error. The plots show the average error along with shaded error bands representing one standard deviation from the mean, allowing us to evaluate both accuracy and variability.

For call options (Fig.~\ref{fig:black_scholes_CALL_results}) presents the relative error for call options across different strike prices. The VQC and DNN produce similar average results, but the variance of the VQC errors is consistently lower. This indicates that the VQC solver is more stable than the DNN when applied to call option pricing.

For put options (Fig.~\ref{fig:black_scholes_PUT_results}) shows the results for put options. In this case, the VQC outperforms the DNN in both average accuracy and error variance. The advantage is especially clear for out-of-the-money puts, where the option value is more sensitive to input changes and the pricing problem becomes more challenging. In these regions, the VQC solver provides more reliable and accurate estimates.

Overall, the VQC demonstrates more stable performance across the portfolio and delivers more accurate results than the DNN in pricing put options. These findings highlight the effectiveness of VQC-based methods in solving BSDEs for financial applications.

\subsection{Hamilton-Jacobi-Bellman Equation}

\subsubsection{PDE Formulation}

We now consider the following nonlinear parabolic PDE of HJB type:
\begin{equation}
\label{eq:hjb_pde}
\frac{\partial u}{\partial t}(t,x) + \Delta u(t,x) - \lambda \| \nabla u(t,x) \|^2 = 0, \quad (t,x) \in [0,T) \times \mathbb{R}^d,
\end{equation}
with terminal condition:
\begin{equation}
\label{eq:hjb_terminal}
u(T,x) = g(x).
\end{equation}

where \( \Delta u(t, x) = \sum_{i=1}^d \frac{\partial^2 u(t, x)}{\partial x_i^2} \) is the Laplacian, which arises from the second-order operator \( \mathcal{L} u = \frac{1}{2} \operatorname{Tr}(\sigma \sigma^\top \nabla^2 u) \) when \( \sigma = \sqrt{2} I \).

Here, \( \lambda > 0 \) is a parameter controlling the strength of the optimal control in the underlying stochastic control problem. As \( \lambda \) increases, the nonlinearity of the PDE also increases.

Following~\cite{han2018solving}, we choose the terminal condition:
\begin{equation} \label{eq:hjb_terminal_g}
g(x) = \log \left( \frac{1 + \|x\|^2}{2} \right).
\end{equation}

\subsubsection{BSDE Representation}

We consider the forward SDE:
\begin{equation}
\label{eq:hjb_forward_sde}
dX_t = \sqrt{2} \, dW_t, \quad X_0 = \xi \in \mathbb{R}^d.
\end{equation}

The corresponding BSDE is:
\begin{equation}\label{eq:hjb_bsde}
\begin{aligned}
dY_t &= \lambda \| Z_t \|^2 \, dt + Z_t^\top \, dW_t, \quad t \in [0,T), \\
Y_T &= g(X_T).
\end{aligned}
\end{equation}

\subsubsection{Exact Solution}

The HJB equation~\eqref{eq:hjb_pde} admits the following exact solution (cf.~\cite{han2018solving}, Eq.~(14)):
\begin{equation}
\label{eq:hjb_exact_solution}
u(t,x) = -\frac{1}{\lambda} \ln \left( \mathbb{E} \left[ \exp \left( -\lambda g\left( x + \sqrt{2}(W_T - W_t) \right) \right) \right] \right).
\end{equation}

This formula allows us to compute a reference solution by Monte Carlo simulation of the Brownian increment \( W_T - W_t \), enabling quantitative assessment of the approximation accuracy.

\subsubsection{Numerical Setup}

We consider dimension \( d=100 \), time horizon \( T=1 \), and discretize the time interval with \( N = 10 \) time steps. We use \( 10,000 \) Monte Carlo sample paths.

We evaluate the solver's performance across a family of HJB problems indexed by \( \lambda \in \{1,2,\dots,20,30,40,50,60\} \). As \( \lambda \) increases, the PDE becomes more nonlinear and the control term dominates.

The architectures are the same as in the Black--Scholes experiment:

\begin{itemize}
    \item DNN: 4 hidden layers, 64 neurons per layer, ReLU activation.
    \item VQC: 2 qubits, 2 VQC layers, batch size 100, learning rate 0.01, 10 epochs.
\end{itemize}

The training objective is to minimize the expected terminal squared loss:
\[
L(\theta) = \mathbb{E} \left[ \left| Y^\theta_T - g(X_T) \right|^2 \right].
\]

\subsubsection{Results and Discussion}

\begin{figure}[h]
    \centering
\includegraphics[width=1\linewidth]{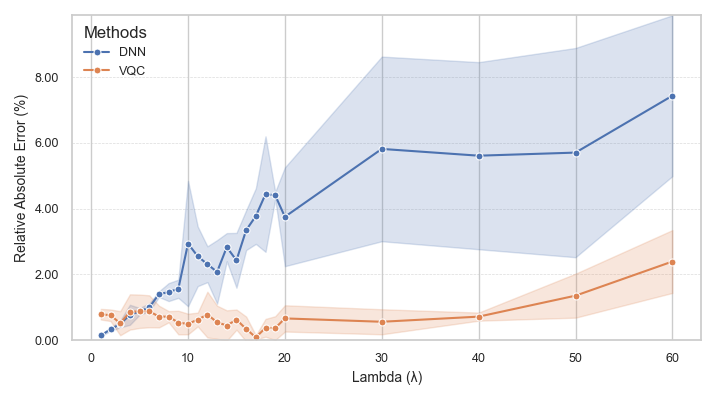} 
    \caption{Relative error (\%) for VQC and DNN across \( \lambda \) values in the HJB equation.}
    \label{fig:hjb_results}
\end{figure}

Figure~\ref{fig:hjb_results} shows the relative error (\%) of both models across the range of \( \lambda \) values.

To assess solver robustness in the nonlinear case, we evaluate both solvers on the HJB equation. Each experiment is repeated five times, and we report the mean and standard deviation of the relative error for different values of the control parameter 
$\lambda$. The plots in Figure~4 show the average error along with shaded error bands representing one standard deviation from the mean, allowing us to evaluate both accuracy and variability.

As $\lambda$ increases, the PDE becomes more nonlinear due to the growing influence of the term $\lambda \|\nabla u\|^2$, which begins to dominate the dynamics of the solution. This introduces additional complexity, making the approximation problem more difficult.

Across all runs, the VQC-based solver consistently achieves lower average error and smaller variance compared to the DNN. The DNN's performance tends to degrade as $\lambda$ increases, with noticeably higher variability, while the VQC maintains stable and accurate approximations even in highly nonlinear regimes.

These results demonstrate that VQCs offer improved stability and robustness in solving nonlinear stochastic control problems. The consistent performance of VQC across repeated trials suggests that it is less sensitive to initialization and training noise, making it a strong candidate for solving high-dimensional PDEs in stochastic control problems.

\section{Conclusion}

In this work, we proposed a quantum machine learning framework for solving high-dimensional parabolic PDEs via their representation as BSDEs. Specifically, we developed a pure VQC-based BSDE solver, where all trainable components reside in the VQC, and classical layers (used for dimensional adaptation) are fixed and randomly initialized.

To evaluate the effectiveness of our approach, we benchmarked the VQC-based BSDE solver against a classical DNN baseline on two canonical problems: the Black–Scholes equation and the nonlinear HJB equation, both in 100 dimensions. The VQC solver consistently achieved lower variance and improved accuracy in most cases, particularly in more challenging regimes such as out-of-the-money options and nonlinear stochastic control scenarios. These results demonstrate that VQCs offer enhanced stability as well as competitive, and often superior, accuracy in solving high-dimensional PDEs.

In future work, we plan to extend this framework to address more complex challenges in quantitative finance. These include pricing financial derivative products with high-dimensional, correlated underlying assets, path-dependent features, and optimal stopping components such as structured interest rate derivatives. We also aim to explore applications in counterparty credit risk, particularly in the computation of X-value adjustments (XVAs). These directions represent important steps toward applying quantum BSDE solvers to realistic, high-impact problems in the financial markets.

\section*{Acknowledgment}
H.~H.~Tseng is supported by the Laboratory Directed Research and Development Program \#24-061 of Brookhaven National Laboratory and National Quantum Information Science Research Centers, Co-design Center for Quantum Advantage (C2QA) under Contract No. DE-SC0012704. The research used resources of the NERSC, under Contract No. DE-AC02-05CH11231 using NERSC award HEPERCAP0033786.

\bibliographystyle{IEEEtran}
\bibliography{refs}

\end{document}